\documentstyle[]{article}
\begin{document}
\begin{center}
{ \huge \bf
On Static Dielectric Response of Microcomposites of the Type
Ferrolastics-Dielectrics For Application in Solid Oxide Fuel
Cells (SOFC). }
\end{center}
\vskip1cm
\begin{center}
{\Large   M. Hudak{\footnote{\large Stierova 23, Kosice}}
\vskip0.5cm
O. Hudak{\footnote{\large Department of Aerodynamics and
Simulations, Faculty of Aviation, Technical University Kosice,
Kosice} }
 }
\end{center}

\newpage
\section*{Abstract}

We describe the static dielectric response of
ferroelastic-dielectric
microcomposites. Dependence on temperature, pressure and
concentration is considered for temperatures in the paraelastic
transition.
In recent years there has been considerable interest in
perovskite oxides for application in solid oxide fuel cells
(SOFC), exhaust gas sensors in automobiles, membranes for
separation processes and as catalysts, which are of the
ferroeleastic nature, f.e. of the ($ LaSrCoO_{3}$) type.
Ferroelastic-dielectric microcomposites and measurement of their
properties may improve properties of perovskite oxides for these
applications.

\newpage
\section{Introduction}

Dielectric mixtures are, \cite{A01}, composites which are made up of
at least two constituents. The outstanding mechanical
properties of many composites, and namely the unique combination
of low density with high strength and stiffness led to extensive
research and to a highly developed technologies \cite{A02},
\cite{A03} and \cite{A04}. In \cite{A01} attention was given to
their physical properties, which affected their use in
electrical applications \cite{A010}, \cite{A011}, \cite{A012}
and \cite{A013}. In recent years there has been considerable
interest in perovskite oxides for applications in solid oxide
fuel cells (SOFC), exhaust gas sensors in automobiles, membranes
for separation processes and as catalysts \cite{A05}.
Requirements for the above materials are high electronic and
ionic conductivity, stability, and satisfactory mechanical
properties \cite{A06}. $ LaCoO_{3}$ based materials are
interesting for these applications due to their high electronic
and ionic conductivity when La is substituted by divalent
cations. $LaCoO_{3}$ has a perovskite-type structure with La on
A-site and Co on B-site, and the oxygen stoichiometry varies
with $O_{2}$ pressure and with temperature. In \cite{A06}
mechanical properties of $LaCoO_{3}$ based ceramics were
studied. The mechanical behavior of lanthanum-cobaltite based
perovskites with a mixed ion-electron conductivity at different
temperatures was studied in \cite{A07}, it is noted that these
materials are environment-friendly. In they there are realized
the direct conversion of fuel energy to electrical energy using
ceramic fuel cells materials and the ion-mediated separation of
oxygen from air materials. Ceramics for the above should be
efficient engineering component exposed to thermal and
mechanical loading conditions. They must simultaneously play the
role of a structural material. Perovskites of the type
$LaCoO_{3}$ \cite{A08} are candidate materials for those
purposes. So they are promising for fabricating membranes in the
high-temperature separation of oxygen from air, the cathodes of
ceramic fuel cells, etc. \cite{A09}. Ferroelastic perovskites
can undergo an energy-absorbing switching rotation of
ferroelastic domains \cite{A10} . The range of papers studying
the mechanical behavior of perovskites is rather modest
\cite{A10}, for example moduli of elasticity \cite{A11},
\cite{A12} strength, \cite{A13} crack resistance and other
characteristics \cite{A14} and \cite{A15} have been considered.
$LaCoO_{3}$ as a promising cathode material for solid oxide fuel
cells with a perovskite type structure has the ability to
reduce itself reversibly at moderate partial pressures of
oxygen, thus producing large numbers of oxygen ion vacancies.
The ion vacancies allow the transport of oxygen ion and cause
oxygen surface exchange, raise mechanical stress by causing
volume expansion of the lattice \cite{A16}. Influence of
temperature change on the properties of $ LaSrO_{3}$, type
LSC-82, has been reported in \cite{A17}. Effect of temperature
on magnetic susceptibility has been reported \cite{A18}.

In these materials behind mechanical properties also electric
properties are important in relation to their mechanical
properties. They are
described by the dielectric susceptibility which may be studied
experimentally To improve properties of $ LaCoO_{3}$ based
materials one may consider the ferroeleastic-dielectric
composites, in which $ LaCoO_{3}$ type material is
ferroeleastic. Recently we have studied dielectric response of
ferroeleastic-dielectric composites for low and for high
concentrations of the ferroelastic material \cite{HS1} and
\cite{HS2}. The dielectric response of ferroelectric micro- and
nano- particles and ferroelectric-dielectric type composites
was studied in \cite{HRP} - \cite{OH2}. When a coupling of the
elastic
strain to the electric polarization is present then the
dielectric
response of a ferroelastic material may be studied, see in
\cite{jona} - \cite{LG}. Also the dielectric response of
ferroelastic-dielectric type microcomposites
may be studied. Response of minerals to changing hydrostatic
pressure p and
temperature T is also interesting property of these materials.
For
example materials of the perovskite type ($LaAlO_{3}$,
$CaAlO_{3}$,
$SrAlO_{3}$, $BaTiO_{3}$, $PbNiO_{3}$, $Pb(Zr,Ti)O_{3}$, ...)
undergo a phase transition from a cubic phase to a phase with
lower
symmetry at some critical temperature. While a study of
their elastic properties is usually done in literature, we will
consider here the dielectric response of materials of
ferroeleastic-dielectric type which is due
to coupling between the elastic strain tensor and the
polarization
Mechanical analysis is usually done at low
frequencies (0.1Hz - 10 Hz) but also measurements at higher
frequencies
are done. Ferroelastic domain wall structure, twinning and other similar phenomena are studied, see in \cite{STKSS} and
\cite{KSSHSS}. Elastic response function (compliance)
shows in Cole-Cole diagrams circular and non-circular behavior
of
these materials in their crystalline and ceramic form
\cite{HRS}. In the
second case multirelaxation phenomena exist in these materials
under
some conditions. Under higher electric and mechanical loading
nonlinear behavior is exhibited by ferroelectric and
ferroelastic
ceramics \cite{E}. Coupling of the elastic strain and electric
polarization does exist at these materials and thus dielectric
response
depends on mechanical forces acting on the composite of the
ferroelastic-dielectric type. This response enables us to study
properties of these microcomposites using dielectric
measurements for
microcomposites under mechanical forces. It is known that
constraint
due to neighboring material lead below the critical temperature
for
transition from paraelastic to ferroelastic phase and due to
shape-change to several forms of the low-temperature phase
\cite{JCD}.  As it is noted by these authors only in very
small grains there exists a single variant of this form. A
strain of
$10^{-4}$ in a grain of size 10 $\mu$m (typical values) is an
order of
magnitude larger to be accommodated in a small displacements of
atoms nearby and at surface. We assume
in our paper that there are small mechanical fields of the order$10^{-3}$ and that particles are with their diameter of the
order of
1$\mu$m. They are microcomposites.  In ferroelastic phase long-range anisotropic
forces may appear
\cite{LSRSB}. In our paper we discuss properties of the
microcomposite
in the paraelastic phase, thus these long-range forces can be
neglected. For solid oxide fuel cells namely this later limit is
used. Small amount of dielectric material in ferroeleastic phase
(microcomposite) may improve their properties.

The aim of this paper is to study the static dielectric response
of microcomposites of the type: ferroelastics-dielectrics in dependence on pressure and temperature. Both
pressure and temperature play role in solid oxide fuel cells.
Measuring dielectric susceptibility of ferroelastic material and
of microcomposites of the type ferroeleastic-dielectric type one
may use to change their properties.
In the second chapter a model of ferroelastic-dielectric
microcomposites is presented. We study the dielectric response
of
ferroelastic particles. Ferroelastic particles and their
dielectric response to static
electric field in the paraelastic limit and as a function of
hydrostatic pressure is studied in the next chapter.
Effective Medium Approximation (EMA) is generally used for
dielectric response description of microcomposites of the type:
ferroelastics-dielectrics, and then studied in low
concentration limit of the dielectric material. In the last
chapter we summarize our results as concerning temperature,
hydrostatic pressure and concentration dependence of the
dielectric material as concerning response of
microcomposites of the type: ferroelastics-dielectrics.

\section{Model of Ferroelastic-Dielectric Microcomposites}

Microcomposites are composited from small
particles.  One of such examples are
ferroelectric-dielectric microcomposites
\cite{HRP} - \cite{OH2}. Ferroelectric properties of particles
may appear in
them due
to presence of the polarization as a primary order parameter. An interesting possibility is to consider materials in which
ferroelectricity is induced as a secondary order parameter.
Primary
order parameter may be the corresponding component of the
elastic
strain tensor. The ferroelectric state is present in such a
material due to a coupling between the
elastic strain tensor and the polarization We have two types of particles in the microcomposite: ferroelastic and dielectric.
Changing
the concentration of these two types of particles in the
microcomposite the response to external fields changes. This
holds for
dielectric response and for other type of responses. We are
calculating
in this paper the dielectric response of such a microcomposite.
For simplicity we consider all particles in
microcomposite of the same diameter d. 
In real materials there exists distribution of
diameters of particles and shapes of particles. In our model we
neglect this
distribution for simplicity.

\section{Ferroelastic Particles: Dielectric Response}

The dielectric response of a ferroelastic particle will be
studied
using a
Landau free energy expansion. We will assume cubic symmetry of
the
ferroelastic material in the paraelastic phase for simplicity. 
For materials of other symmetry
the approach is similar. In ferroelastic materials the primary
order
parameter is an elastic strain tensor. Secondary order parameter
is
polarization.
While the elastic strain tensor is coupled to the external
mechanical fields
(hydrostatic pressure, uniaxial stress, shear stress), the
secondary order
parameter is coupled to an external electric field. We will use
in our
calculation time and space dependent external fields in general.
However for
microcomposites the quasistatic approximation for dielectric
response is a convenient
approximation because the wavelength of the electric field is
usually much
larger than the diameter of the particle. The space dependence
of
the external electric field may be neglected.

To find the free energy expansion we have to find invariants of
the
primary order parameter and of the secondary order parameter,
and of
primary and secondary order parameter coupled.
In cubic materials there are the following invariants from the
components of the elastic strain tensor, the primary
order parameter, $\epsilon_{i,j}$ where $i,j = 1,2,3$ denotes
axis of
the cubic material (Einstein sum rule is used)

\begin{equation}
\label{1}
\epsilon_{i,j} \epsilon_{j,i} = 3 \epsilon^{2} + 6  \phi^{2}
\end{equation}

which is of the second order in the elastic strain  tensor, and

\begin{equation}
\label{2}
\epsilon_{i,j} \epsilon_{j,k} \epsilon_{k,i} = 3 \epsilon
 (\epsilon^{2} + 2  \phi^{2}) + 6 (2 \epsilon 
\phi + \phi^{2})\phi .
\end{equation}

This term is of the third order in the elastic strain tensor. 
Here we denoted on-diagonal terms in $ \epsilon_{k,i} $ as $
\epsilon $ and off-diagonal
terms of the elastic strain tensor $ \epsilon_{k,i} $ as $ \phi
$ for the cubic phase.

The coupling between the primary order parameter (elastic
strain tensor) and the
secondary order parameter (electric polarization) is of the
first
order in the elastic strain tensor and of the
second order in the polarization. We will consider in the free
energy expansion only these two invariants
for the elastic strain tensor described above, (\ref{1}) and
(\ref{2}).
Thus in the corresponding to polarization part of the free
energy
expansion there will be the second order term, the fourth order
terms and the sixth order
terms in general.

The coupling between the elastic strain tensor $\epsilon_{k, i}$
and the polarization vector $ P_{i} $ has the form
 
\begin{equation}
\label{3}
\epsilon_{i,j} P_{i}  P_{j} = \epsilon P^{2}.
\end{equation}

The polarization vector $P_{i}$ is assumed to have a nonzero
component only in the
x-direction, we assume that external electric field will be
applied in
this direction.
Then the free energy $F$ expansion has the form

\begin{equation}
\label{4}
F = \int dV [\frac{B}{2} ( \epsilon^{2} + 2 \phi^{2} ) +
\frac{C}{3}(
 \epsilon
 (\epsilon^{2} + 2  \phi^{2}) + 2  (2  \epsilon 
\phi + \phi^{2})\phi ) + \Gamma P^{2} \epsilon +
\frac{\alpha}{2}
 P^{2} + \frac{\beta}{4} P^{4} +  
\end{equation}
\[+ \frac{\gamma}{6}
 P^{6} - E.P - \epsilon_{i,j}  \sigma_{j,i}]. \]

The last term in the free energy expansion depends on the stress tensor This tensor may be hydrostatic pressure p, then $
\sigma_{j,i}=
\delta_{j,i}p$, uniaxial stress $ \sigma_{x,x}= \sigma $ or
shear
stress $ \sigma_{x,y} $. The constants in the free energy
expansion
(\ref{4}) are positive and temperature independent, with the
exception
of the constant $B$, for which $ B = B_{0}(T - T_{c})$ where
$B_{0}$
is a constant, $T_{c}$ is a critical temperature for the
transition from
the paraelastic to the ferroelastic phase.
The expansion constants in (\ref{4}) may be in fact temperature
and
hydrostatic pressure dependent. We will not consider this
dependence with the exception of the constant B. 
Depending on the field applied (electric, mechanical) we
calculate the
response of the material described by the free energy expansion
(\ref{4}). We will assume that the surface charge is compensated
in the case of
polarized particles, and surface effects are neglected for the strain.

Let us now consider the paraelastic phase.
In this case the free energy expansion from (\ref{4}) takes the
form in which second order and fourth order terms in the
polarization are taken into
account and the first order terms in the elastic strain tensor
are taken into
account. Note that the second order in the electric polarization
is
corresponding to
the first order in the elastic strain tensor.
However the coupling term is of the fourth order in
polarization, so
also the second order term in elastic strain tensor is taken
into account.
Then the free energy $F$ expansion has the form

\begin{equation}
\label{5}
F = \int dV [\frac{B}{2} ( \epsilon^{2} + 2 \phi^{2} ) + \Gamma
P^{2} \epsilon + \frac{\alpha}{2}
P^{2} + \frac{\beta}{4} P^{4} - E.P - \epsilon_{i,j}
\sigma_{j,i}].
\end{equation}

Let us consider the static case of the electric field and of the hydrostatic pressure in the following section.

\section{Dielectric Response of Ferroelastic Particles: Static
High
  Temperature Limit and Hydrostatic Pressure Dependence}

In this case the free energy  expansion (\ref{5}) has the form:
 
\begin{equation}
\label{6}
F = \int dV [\frac{B}{2} ( \epsilon^{2} + 2 \phi^{2} ) + \Gamma
P^{2} \epsilon + \frac{\alpha}{2}
 P^{2} + \frac{\beta}{4}
 P^{4} - E.P - 3 \epsilon p] .
\end{equation}

Here p is the hydrostatic pressure. The Lagrange-Euler equations
for the most stable state at a given
electric field and hydrostatic pressure have the form

\begin{equation}
\label{6.1}
B \epsilon - 3p + \Gamma P^{2}= 0
\end{equation}
\[ \alpha P + \beta P^{3} + 2 \Gamma \epsilon P - E = 0 \]
\[  \phi = 0 .\]

We obtain that the dielectric permitivity $\epsilon_{f}$ has the
form

\begin{equation}
\label{6.2}
\epsilon_{f} = \frac{1}{\alpha^{*}} \equiv \frac{1}{\alpha}
(\frac{1}{1 + \frac{ 6 \Gamma p}{B \alpha}
+(\frac{\beta}{\alpha} - \frac{2 \Gamma^{2}}{B \alpha}) P^{2}})
\end{equation}

When the condition

\begin{equation}
\label{6.3}
(\alpha + \frac{6\Gamma p}{B})^{2} >> \left| \beta - \frac{2
\Gamma^{2} }{B} \right| E^{2}
\end{equation}

is fulfilled, e.i. for small electric field $E^{2} <<
\frac{(\alpha + \frac{6\Gamma p}{B})^{2}}{\left| \beta - \frac{2
\Gamma^{2} }{B} \right| })$ for given temperature, pressure p and
constants $\alpha$, $\beta$, $\Gamma$, $B_{0}$ and $T_{c}$, then
the dielectric permitivity $\epsilon_{f}$ has the form

\begin{equation}
\label{6.23}
\epsilon_{f} = \frac{1}{\alpha^{*}} \equiv \frac{1}{\alpha}
(\frac{1}{1 + \frac{ 6 \Gamma p}{B \alpha}} )
\end{equation}

The dielectric constant $\frac{1}{\alpha}$ is $(\frac{1}{1 +
\frac{ 6 \Gamma  p}{B \alpha}} ) $ times changed than that in
zero pressure case. We assume that $B > 0$ and that $B = B_{0}(T - T_{c})$. Taking higher harmonics $\frac{\beta}{4}
P^{4}$ in the free energy expansion into account we obtain that
the dielectric permitivity $ \epsilon_{f}^{*} \equiv
\frac{1}{\alpha^{**}} $ has approximately the form

\begin{equation}
\label{6.21}
\epsilon_{f}^{*} = \frac{1}{\alpha^{**}} \equiv \frac{1}{\alpha}
\frac{1}{(1 + \frac{ 6 \Gamma p}{B \alpha} + (\frac{\beta}{
\alpha} -\frac{2 \Gamma^{2} }{B \alpha})(\frac{1}{\alpha^{*}}E)^{2})}
\end{equation}

Here we have taking $ P \approx \frac{1}{\alpha^{*}}E$ in
(\ref{6.2}). The therm $P^{2}$ in (\ref{6.21}) may be neglected
for small fields E and then we obtain from (\ref{6.21}) the form
(\ref{6.2}).
As we can see from the equation (\ref{6.21}) the dielectric
permitivity may behave differently now with increasing
hydrostatic pressure, and is electric field $E$ dependent.

\section{Effective Medium Approximation-General Formulation
for Dielectric Response of Composites of the Type: Ferroelastics-Dielectrics }

The effective permitivity $\epsilon_{eff} $ may be obtained,
\cite{HRP}, from the
effective medium approximation for the whole interval of
concentrations $ 0 \leq x \leq 1 $ of the dielectric components
of the ferroelastic-dielectric microcomposite

\begin{equation}
\label{9}
x  \frac{\epsilon_{d} - \epsilon_{eff} }{\epsilon_{d} +2 
\epsilon_{eff} } + (1-x) \frac{\epsilon_{f} - \epsilon_{eff}
}{\epsilon_{f} +2 \epsilon_{eff}} = 0
\end{equation}

This approximation is based on the response of a spherical
particle to
the whole effective microcomposite. Both components of the
microcomposite are taken into
account symmetrically.  For the permitivity of the
hard dielectric
material we take the constant dielectric permitivity
$\epsilon_{d}$ and for the permitivity of the ferroelastic
material $\frac{1}{\alpha}\frac{1}{(1 + \frac{ 6 \Gamma p}{\alpha B })}$.
The effective permitivity calculated from (\ref{9}) will
describe dielectric response of the
microcomposite on the electric field and on the hydrostatic
pressure. Let us now
consider the limiting case: the limit of small concentration of
the hard dielectric material.

\section{High Concentration of the Ferroelastic Material}

For low concentration of dielectric particles in the
ferroelastic matrix
 we may calculate the dielectric response of the
microcomposite from (\ref{9}). In this case we have $ x $ near
the value 0.
The value $x=0$ corresponds to ferroelastic material only.
Thus $ x $ is a small parameter now . The effective dielectric
permitivity $ \epsilon_{eff}$ is given as

\begin{equation}
\label{12}
\epsilon_{eff} = \epsilon_{f} + 3 x \epsilon_{f} 
\frac{\epsilon_{d} - \epsilon_{f}}{\epsilon_{d} + 2 \epsilon_{f}
}
\end{equation}

Substituting for the ferroelastic permitivity from (\ref{6.2}),
where we neglect the term $P^{2}$, we
obtain hydrostatic pressure, temperature and concentration
dependence of the
effective dielectric permitivity in the cubic phase of
ferroelastic matrix with dielectric material in this limit as

\begin{equation}
\label{13}
\epsilon_{eff} = \frac{1}{\alpha}\frac{1}{(1 + \frac{ 6 \Gamma
p}{ B \alpha })} +
\end{equation}
\[ + 3 x \frac{1}{\alpha}\frac{1}{(1 + \frac{ 6 \Gamma p}{B
\alpha })}.
\frac{\epsilon_{d} - \frac{1}{\alpha}\frac{1}{(1 + \frac{ 6
\Gamma p}{B \alpha})}}{
\epsilon_{d} + 2 \frac{1}{\alpha}\frac{1}{(1 + \frac{ 6 \Gamma
p}{ B \alpha})}} \]

As we can see for dielectric material with $ \epsilon_{d}$ such
that $(\epsilon_{d} - \frac{1}{\alpha}\frac{1}{(1 + \frac{ 6
\Gamma \alpha p}{B})}) >0 $ we may increase the effective
dielectric response of the microcomposite by the hard dielectric
material. This may influence oxygen and electron particles in
their move in the material.

The response of the diagonal component $\epsilon_{elastic, f}$
of the elastic strain tensor, neglecting the term $P^{2}$, has
the form of the Hook law

\begin{equation}
\label{6.11}
\epsilon_{elastic, f} = \frac{ 3p}{B}
\end{equation}

for the ferroelastic material in the paraelastic phase. For the microcomposite in this
limit we obtain

\begin{equation}
\label{12,1}
\epsilon_{elastic, eff} = \frac{ 3p}{B} + 3 x \frac{ 3p}{B}
\frac{\epsilon_{elastic, d} - \frac{ 3p}{B}}{\epsilon_{elastic,
d} + 2 \frac{ 3p}{B} } .
\end{equation}

Here $\epsilon_{elastic, d}$ is the diagonal component of the
elastic strain tensor of the dielectric material. For $p_{c} =
\epsilon_{elastic, d} \frac{B}{3}$ the contribution of the
dielectric component in the microcomposite to the response
$\epsilon_{elastic, eff}$ changes the sign for nonzero
concentration x as can be seen from

\begin{equation}
\label{12,11}
\epsilon_{elastic, eff} = \frac{ 3p}{B} + 3 x \frac{ 3p}{B}
\frac{\frac{p_{c}}{p} -  1}{\frac{p_{c}}{p} + 2 } .
\end{equation}

\section{Summary}

We studied here the static dielectric response of
microcomposites:
ferroelastic-dielectric. A model for such a microcomposite was formulated. Dielectric properties of ferroelastic particles were studied 
We
considered a coupling of the elastic strain tensor to the
electric
polarization. While the primary order parameter is coupled to
external mechanical
fields, the secondary order parameter is coupled to external
electric
field. We have found the free energy expansion using
invariants of the primary order parameter, of the secondary
order
parameter, and of their mixed terms. We assumed that in the free energy expansion only the coefficient B of the second order of
the
primary order parameter is temperature dependent, and that other parameters are temperature independent. All of these
coefficients are
assumed to be hydrostatic pressure independent. Surface charges
are assumed to
compensate the dipole moment in ferroelastic particles in which
an
electric dipole appears. We apply hydrostatic pressure on
the microcomposite and consider properties of such a
microcomposite here.  The most important role in the free
energy expansion plays terms of the second order in the elastic
strain tensor
and of the second order in electric polarization The fourth
order
terms of the polarization are small for small electric fields
and we do not consider them here.
We studied the dielectric response of the ferroelastic particles for paraelastic phase  and its hydrostatic pressure dependence.

The dielectric response of a particle is depending on
hydrostatic pressure:
increasing hydrostatic pressure the dielectric response is
smaller for the constant $\Gamma$
positive and larger for the constant $\Gamma$
negative (we assume $(1+\frac{6 \Gamma p}{B \alpha})>0$).

The effective medium approximation theory is
formulated for study of the dielectric response of the
microcomposite. The limit of small concentrations of the
hard dielectric material is studied here and the effective
dielectric constant is
calculated.
Increasing temperature the effective dielectric
constant of the microcomposite the dielectric
response
of the microcomposite is tending to the dielectric
response
of the microcomposite which is under no hydrostatic pressure.

The low concentration of the dielectric material increases the
effective permitivity of the microcomposite with nonzero hydrostatic pressure for
dielectric material with $ \epsilon_{d}$ such
that $(\epsilon_{d} - \frac{1}{\alpha}\frac{1}{(1 + \frac{ 6
\Gamma \alpha p}{B})}) >0 $.

In our model above we did not consider the mechanical
inclusion/matrix interactions.  We
discuss temperatures in which there is the paraelastic phase.
Effect of elastic clamping
was not considered here. For its discussion for an improper and a pseudoproper ferroelastic inclusion see \cite{PS}. For
ellipsoidal shapes of the ferroelastic inclusions the order
parameter and
 strain are uniform inside the inclusion. For improper and
 pseudoproper ferroelastic inclusions and polycrystalline
inclusion/matrix interaction renormalizes the constant of the
Landau
free energy expansion of the order parameter. For proper
ferroelastic
materials which we consider here 3D clamping of the crystal
inclusion
 in the matrix is not considered, we assume the mechanical
equilibrium of the inclusion/matrix system is present at
temperatures
considered.
As concerning crystal structure of perovskite
$La_{1-x}Sr_{x}CoO_{3}$ for $ 0.0 < x < 0.7 $ it was studied by
\cite{MK}. The space group was assigned to rhombohedral
$R3^{-}c$ in the range $ 0.0 < x < 0.5 $ and to cubic $Pm3m$ in
the range $0.55 < x $.

\end{document}